\documentclass[epj]{svjour_pb}
\usepackage[dvips]{hyperref}
\usepackage{times}
\usepackage{graphics}
\usepackage{epsfig}
\usepackage[latin1]{inputenc}
\usepackage[english]{babel}
\usepackage{array}
\usepackage{amssymb}
\usepackage{amsmath}
\newcommand{\ee}{\mbox{${\mathrm{e}}^+ {\mathrm{e}}^-$}}

\newcommand{\mhmax}      {\mbox{$m_{\mathrm{h}}{\mathrm{-max}}$}}

\newcommand{\bb}         {\ensuremath{\mathrm{b}\bar{\mathrm{b}}}}

\newcommand{\mHone}         {\mbox{$m_{\mathrm{H}_{1}}$}}
\newcommand{\mHtwo}         {\mbox{$m_{\mathrm{H}_{2}}$}}

\newcommand{\mh}         {\mbox{$m_{\mathrm{h}}$}}
\newcommand{\mA}         {\mbox{$m_{\mathrm{A}}$}}

\newcommand{\Hone}{\ensuremath{\mathrm{H}_1}}
\newcommand{\Htwo}{\ensuremath{\mathrm{H}_2}}

\newcommand {\Ao}        {\ensuremath{\mathrm{A}}}
\newcommand {\ho}        {\ensuremath{\mathrm{h}}}
\newcommand {\Zo}        {\ensuremath{\mathrm{Z}}}

\newcommand{\tanb}       {\mbox{$\tan\beta$}}

\newcommand{\ra}        {\mbox{$\rightarrow$}}   

\begin{document}

\title{Searches for Neutral Higgs Boson and Interpretations in the MSSM at LEP}
\author{Philip~Bechtle\thanks{Speaker; on behalf of the LEP collaborations}  
}                     
%
%
\institute{DESY, Notkestr.~85, 22607~Hamburg, Germany; E-mail: \texttt{philip.bechtle@desy.de}}
\date{Received: 1 Oct 2003 / Accepted: 20 Nov 2003 / \\
Published Online: 26 Nov 2003 -- \copyright{} Springer-Verlag / Societ\`a Italiana di Fisica 2003}
%
\abstract{
This paper discusses recent publications of the LEP collaborations DELPHI, L3 and OPAL on searches
for Higgs bosons motivated by MSSM scenarios as well as their interpretation in the MSSM.
With the final publication of the LEP collaborations available 
or awaited, more and more interpretations in different MSSM models, including both CP 
conserving and CP violating, become available. Also specialized analyses close open areas
in the parameter space. In the same time, better theoretical calculations with an increased 
maximal mass of the $\ho$ boson were presented. Both the new scenarios as well as the new theoretical 
limit on $\mh$ has consequences for the limits from LEP.
The searches, the models in which they are interpreted and the 
implications of the LEP results for future SUSY searches, especially on the $\tanb$ limit, 
are presented here.  
} 
\authorrunning{P. Bechtle}
\maketitle
%

%
%
\section{Introduction}\label{sect:intro}

In the Standard Model (SM) it is generally assumed that the Higgs mechanism is resposible for the
breaking of electroweak symmetry and for the generation of elementary
particle masses.
The Minimal Supersymmetric Standard Model (MSSM) is the SUSY extension of the
SM with minimal new particle content. It introduces two complex Higgs field
doublets.
The MSSM predicts five Higgs bosons: three neutral and two charged ones.
At least one of the neutral Higgs bosons is predicted to have its mass close
to the electroweak energy scale, providing a high motivation to the searches
at current and future colliders.


In the MSSM the Higgs potential is assumed to be invariant
under CP transformation at tree level. However, it is possible to break CP
symmetry in the Higgs sector by radiative corrections, especially by
contributions from complex trilinear couplings $A_{\mathrm{t,b}}$ of third generation 
scalar-quarks~\cite{Pilaftsis:1999qt}. 

Since the input parameter space is generally too large
to be scanned completely, so called benchmark scenarios (cf. Tab.~\ref{tab:scenarios}) 
have been proposed~\cite{newbenchmarks,lhc_benchmarks,Carena:2000ks}, 
each emphasising a certain phenomenological situation. The parameters
$\tanb=v_2/v_1$ and $m_{\Ao}$ governing the Higgs sector on tree-level are scanned, while all
parameters on loop level are kept fixed for one scenario. CP conserving (CPC) and 
CP violating (CPV) scenarios exist. 

\begin{table}
  \caption{The MSSM scenarios used by the LEP collaborations, proposed in 
    \cite{newbenchmarks},  \cite{lhc_benchmarks} and \cite{Carena:2000ks}}
  \label{tab:scenarios}       
  \begin{center}
  \begin{tabular}{ll}
    \hline\hline\noalign{\smallskip}
    \multicolumn{2}{c}{CP conserving}\\
    \hline\noalign{\smallskip}
    No Mixing &  No mixing in the stop-sbottom sector \\
    $\mathrm{m}_h$\,max &  Maximum $\mh$ for given $\tanb,\mA$ \\
    Large $\mu$ &  Always kinematically accessible,\\
                &   but $\ho\ra\bb$ suppressed \\
    $\mathrm{m}_h$\,max$^{+}$ &  like $\mathrm{m}_h$\,max, but
    favoured by $(g-2)_{\mu}$.\\
    constr. $\mathrm{m}_h$\,max &  like $\mathrm{m}_h$\,max,
    but favoured by $(\mathrm{b}\ra\mathrm{s}\gamma)$  \\
    gluophobic &  $\ho\mathrm{gg}$ coupling suppressed, \\
               &  reduced LHC production cross section \\
    small $\alpha_{\mathrm{eff}}$ &  $\ho\ra\bb$ suppressed by 
    cancellation of \\
          & $\tilde{\mathrm{b}}-\tilde{\mathrm{g}}$ loops \\
    \hline\noalign{\smallskip}
    \multicolumn{2}{c}{CP violating}\\
    \hline\noalign{\smallskip}
    CPX &  Mixing of CP- and mass-eigensates \\
        &  several derivates under study \\
    \hline\hline
  \end{tabular}
  \end{center}
\end{table}


Depending on the parameters of the MSSM, Higgs Bosons can be produced in Higgstrahlung $\ee\ra\ho\Zo$, as
in the SM, or in pair production $\ee\ra\ho\Ao$.
Flavour independent Higgs decays, 
Higgs decays into invisible particles or decays of the type $\ho\ra\Ao\Ao$ additionally present 
new topologies.

%

Since the end of LEP running,
the LEP collaborations have developed new searches closing some of these 
unexcluded areas in the parameter space, thereby giving important new information about 
the tasks left over for future experiments. This publication will focus on new searches
dedicated to formerly uncovered final states described in Section~\ref{sect:searches},
on the consequences of new theoretical developments in Section~\ref{sect:newbenchmarks} 
and on the interpetation of the MSSM Higgs searches in the benchmark scenarios in
Section~\ref{sect:interpretations}.

%
%
\section{Searches for Higgs bosons in the MSSM}\label{sect:searches}

The search for Higgs bosons in the MSSM uses a large variety of channels.
The SM production channels 
Higgsstrahlung $\ee\ra\ho\Zo$ and Boson fusion are reintrerpreted in the MSSM. Additionally, 
dedicated searches for Higgsstrahlungschannels with Higgs decays in the MSSM exist.
Also, pair production $\ee\ra\ho\Ao$ and Yukawa production $\ee\ra\bb\ho/\Ao$ channels are used. 
New searches comprise the search for $\ee\ra\ho\Zo$ with $\ho\ra\Ao\Ao$, with 
$\mA<10$~GeV below the $\bb$ production threshold~\cite{opal:lowma} by OPAL. The same final state with
heavier $\mA$ has been sought-after for the interpretation in CPV models~\cite{opal:mssmnote}.

The search from DELPHI~\cite{delphi:invis} for invisibly decaying Higgs bosons 
has been interpreted in a modified $\mhmax$ scenario with $M_2=\mu=150$~GeV.
It shows that the benchmark
scenarios from Tab.~\ref{tab:scenarios} do not cover the full range of possible MSSM 
topologies in the Higgs sector, since the search for invisibly decaying Higgs bosons 
is needed to cover areas unexcluded by the standard searches.

%

%
%
\section{New benchmark scenarios and Higgs mass 
  calculations in the MSSM}\label{sect:newbenchmarks}

The calculations of the observables of the Higgs sector, the masses, branching ratios and 
cross-sections, depending on the choice of SUSY parameters 
have been performed using two different calculation tools. 
FEYNHIGGS~\cite{feynhiggs,Heinemeyer:CFeyn} is based on the two-loop diagrammatic approach
of~\cite{MSSMMHBOUND7},
and SUBHPOLE/CPH~\cite{Carena:2000ks} is 
based on the one-loop renormalization-group improved calculation
of~\cite{MSSMMHBOUND5,carenamrennawagner}. 

%
%

The first three CPC benchmark scenarios of Tab.~\ref{tab:scenarios} 
have traditionally been considered in the past and have
now been extended to scenarios 4 and 5, motivated by limits on
the branching ratio of the inclusive decay of a B meson into 
strange particle states and a photon 
$\mathrm{B}\ra\mathrm{X}_{\mathrm{s}}\gamma$ and muon anomalous magnetic moment
$(g-2)_{\mu}$ measurements.
The last two CPC benchmark scans are aiming to set the stage for future
Higgs searches at the LHC. There, some of the dominant search channels
would be suppressed, resulting in a reduced search sensitivity. 
The CPV scenario CPX maximises the mixing of CP- and masseigenstates and
has been tested wit hseveral different parameter settings~\cite{opal:mssmnote}.

\begin{figure}
  \begin{center}
    \epsfig{file=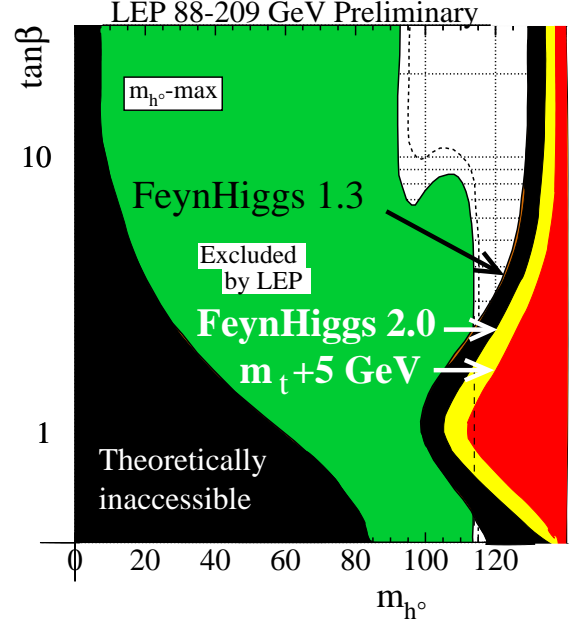, width=0.5\textwidth}
  \end{center}
  \caption{Expected exclusion areas of the $\mhmax$ scenario in the $\mh,\tanb$ projection for increased
    upper bounds on the mass of the $\ho$ boson. The experimental exclusion areas shown 
    are those of the LEP combination 2001~\cite{lephiggs:2001}. The expected theoretical
    upper bounds for $\mh$ given new order $\alpha_{\mathrm{t}}^2$ loop corrections
    to the Higgs mass is shown in light grey (yellow) and the additional effect of a possible shift in $m_{\mathrm{t}}$
    by $+1\sigma$ is overlaid in dark grey (red). The theoretically inaccessible area is forbidden by theory.
  }
  \label{fig:mhmax}       
\end{figure}


%
%

With respect to the calculations used for the MSSM Higgs LEP combination in 
2001~\cite{lephiggs:2001}, new 2-loop calculations of top loop corrections to
the Higgs boson mass have become available~\cite{newcalc}. They shift the maximal
$\mh$ achievable in the $\mhmax$ scenario upwards by up to $5$~GeV. The maximal
$\mh$ lies at about 135~GeV. While the expected experimental lower limit on
$\mh$ for low $\tanb$ is not expected to change much with respect to the latest
LEP exclusion (cf. Fig.~\ref{fig:mhmax}), the theoretical upper limit on $\mh$ 
shifted from the border of the black area to the border of the light grey (yellow) area. 
If the top mass would additionally shift upwards from its current central value of
$m_{\mathrm{t}}=174.3\pm5.1$~GeV~\cite{RPP2000} by only one sigma (which could well be the case, given
latest measurements from D0 in the leptonic decay channel~\cite{d0:newmtop}),
then the upper limit on $\mh$ would increase again by almost 5~GeV, as indicated
by the border of the dark grey (red) area in Fig.~\ref{fig:mhmax}.

This example shows that new theoretical developments and a higher precision on
$m_{\mathrm{t}}$ could well influence the exclusion of low $\tanb$ by LEP. A higher 
precision on  $m_{\mathrm{t}}$ is therefore highly desireable. This could also have
implication on the search channels that have to be investigated at future accelerator 
experiments at LHC searching for a MSSM Higgs boson, where the region of small $\tanb$ 
can not be regarded as excluded by LEP. 

%
%
\section{Interpretation of the Higgs searches in the MSSM}\label{sect:interpretations}

\begin{figure}
  \begin{center}
    \epsfig{file=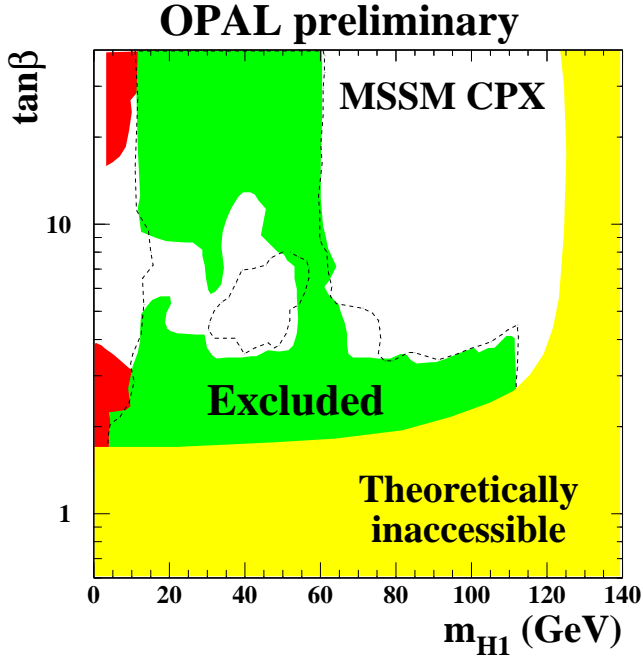, width=0.53\textwidth}
  \end{center}
  \caption{Exclusion area of the CPX scenario in the $\tanb,\mHone$ projection from OPAL. 
    The theoretically inaccessible area is forbidden by theory.}
  \label{fig:cpx}       
\end{figure}

The combination of all Higgs searches of one experiment is used to derive \cite{delphi:mssm,l3:mssm,opal:mssmnote} 
exclusions
in the MSSM parameter space for the scenarios in Tab.~\ref{tab:scenarios}. No major changes
with respect to previous interpretations are recieved for the no-mixing, $\mhmax$ and large-$\mu$
scenarios. The latter now can be almost completely exluded by one experiment alone \cite{delphi:mssm},
thanks to flavour independent searches~\cite{eps:flavindep_proc}.
OPAL has also studied~\cite{opal:mssmnote} the new CPC scenarios 4 to 7 from Tab.~\ref{tab:scenarios}. 

In summary, the parameter choices of the new CPC benchmark scenarios introduce no need for new searches 
at LEP. Latest in a LEP combination, all topologies are covered up to the kinematic limits of the 
production channels. 
The limits on 
$\mh$ and $\mA$ are around 85 to 90~GeV for all CPC scenarios~\cite{aleph:mssm,delphi:mssm,l3:mssm,opal:mssmnote}.

%
%

The CPX scenario with maximal CP violation in the Higgs sector shows a decoupling of the 
lightest Higgs bosons $\Hone$ from the $\Zo$ in the intermediate $\tanb$ range from 4 to 10 ($\Hone$
and $\Htwo$ being the lightest and next-to-lightest Higgs boson mass eigenstates). There 
anyhow $\Htwo$ couples to the $\Zo$ and is heavier than around $100$~GeV. Where kinematically 
accessible, the decay $\Htwo\ra\Hone\Hone$ is dominant. Another difference to the common CPC 
scenarios is the large mass difference $\mHtwo-\mHone$ in the range with dominant pair production.

Fig.~\ref{fig:cpx}~\cite{opal:mssmnote} shows the exclusion areas of the CPX scenario. 
In the region with dominant $\ee\ra\Htwo\Zo\ra\Hone\Hone\Zo$ production at intermediate $\tanb$
and $\mh<50$~GeV open areas emerge. It is expected that a LEP combination will be able to close these holes.
Also the lower limit on $\mHone$ in the large $\tanb$ region, where pair production dominates, is reduced due
to the large $\mHtwo-\mHone$. At $\tanb>5$ and $\mHone<10$~GeV, below the $\bb$ production threshold and
in the pair production region, hardly any experimental constraints exist, since no pair production searches
for $\mHtwo\approx100$~GeV and $\mHone<10$~GeV exist. Only at large $\tanb>20$ Yukawa production searches can
be used.


%
%
\section{Conclusions}\label{sect:conclusions}

The developments in the MSSM Higgs searches at LEP after the end of LEP data taking in November 2000 exhibit
four important lessons. First, also the increased set of CPC and CPV benchmark scenarios do not cover the 
full range of possible experimental phenomena in the MSSM Higgs sector. Therefore, secondly, a large variety of
individual searches is necessary to cover the rich physics spectrum of the MSSM Higgs sector, which only now
become fully available. 

Third, new theoretical developments can influence limits on MSSM parameters. Especially the $\tanb$ exclusion
of the final LEP combination could be affected. This is 
also important for the possible MSSM topologies in Higgs searches at the LHC.
The importance of external measurements like $m_{\mathrm{t}}$ from the Tevatron becomes evident. A greater
precision on $m_{\mathrm{t}}$ would be benefitial.
Fourth, CPV scenarios show that there is still no strict lower limit on the Higgs mass from LEP. Especially
in regions with low $\mHone$, but either dominant $\ee\ra\Htwo\Zo$ or dominant 
$\ee\ra\Hone\Htwo$ production no $\tanb$ independent limit on the Higgs mass exists. Also these regions must 
probably be sought by future colliders.


%
%
%


\begin{thebibliography}{}
%
%

\bibitem{Pilaftsis:1999qt}
A.~Pilaftsis and C.~E.~Wagner,
Nucl.\ Phys.\ B {\bf 553} (1999) 3.


\bibitem{newbenchmarks} 
M.~Carena, S.~Heinemeyer, C.~E.~Wagner and G.~Weiglein,
arXiv:hep-ph/9912223.


\bibitem{lhc_benchmarks} 
M.~Carena, S.~Heinemeyer, C.~E.~Wagner and G.~Weiglein,
Eur.\ Phys.\ J.\ C {\bf 26} (2003) 601.


\bibitem{Carena:2000ks}
M.~Carena, J.~R.~Ellis, A.~Pilaftsis and C.~E.~Wagner,
Phys.\ Lett.\ B {\bf 495} (2000) 155.

\bibitem{opal:lowma}
G.~Abbiendi {\it et al.}  [OPAL Collaboration],
Eur.\ Phys.\ J.\ C {\bf 27} (2003) 483.


\bibitem{opal:mssmnote}
OPAL Collaboration, \emph{Search for Neutral Higgs Bosons Predicted
by CP Conserving and CP Violating MSSM
Scenarios with the OPAL detector at LEP}, 2003, OPAL~PN524

\bibitem{delphi:invis}
[DELPHI Collaboration], \emph{Searches for invisibly decaying Higgs bosons 
  with the DELPHI detector at LEP}, DELPHI~2003-036~CONF~656.


\bibitem{feynhiggs} S.~Heinemeyer, W.~Hollik and~G. Weiglein, 
       Comp.~Phys.~Comm.~{\bf 124} (2000) 76; Also see {\tt http://www.feynhiggs.de}.


\bibitem{Heinemeyer:CFeyn} 
                    M.~Frank, S.~Heinemeyer, W.~Hollik and G.~Weiglein,
                    hep-ph/0212037.\\
                    Also see
                    {\tt www.feynhiggs.de}.

\bibitem{MSSMMHBOUND7} S.~Heinemeyer, W.~Hollik and G.~Weiglein, Eur. Phys. Jour. {\bf C9} (1999) 343.

%

\bibitem{MSSMMHBOUND5} M.~Carena, M.~Quir\'os and C.E.M.~Wagner, Nucl. Phys. {\bf B461} (1996) 407.

\bibitem{carenamrennawagner} M.~Carena, S.~Mrenna and C.~Wagner, Phys. Rev. {\bf D60} (1999) 075010.

%


\bibitem{lephiggs:2001}
ALEPH, DELPHI, L3 and OPAL Collaborations, OPAL~TN699

\bibitem{newcalc}
A.~Brignole, G.~Degrassi, P.~Slavich and F.~Zwirner,
Nucl.\ Phys.\ B {\bf 631} (2002) 195;\\
G.~Degrassi, S.~Heinemeyer, W.~Hollik, P.~Slavich and G.~Weiglein,
Eur.\ Phys.\ J.\ C {\bf 28} (2003) 133.

\bibitem{RPP2000} D.E. Groom {\it et al}, Eur. Phys. J. {\bf C15} (2000) 1,
available on the PDG WWW pages (URL: {\tt http://pdg.lbl.gov/}).

\bibitem{d0:newmtop}
M.~Coca  [CDF \& D0 Collaborations],
FERMILAB-CONF-03-238-E
{\it Presented at Flavor Physics and CP Violation (FPCP 2003), Paris, France, 3-6 Jun 2003}

\bibitem{delphi:mssm}
J.~Fernandez  [DELPHI Collaboration],
DELPHI 2003-045-CONF-665, contributed paper no.~320


\bibitem{l3:mssm}
P.~Achard {\it et al.}  [L3 Collaboration],
Phys.\ Lett.\ B {\bf 545} (2002) 30

\bibitem{eps:flavindep_proc}
M.~Boonekamp, {\it Flavour and Model Independent Higgs Searches}, 
these proceedings


\bibitem{aleph:mssm}
A.~Heister {\it et al.}  [ALEPH Collaboration],
Phys.\ Lett.\ B {\bf 526} (2002) 191.





%
%
\end{thebibliography}
\end{document}